\documentclass{article}

\usepackage[preprint]{neurips_2026}

\usepackage[utf8]{inputenc}
\usepackage[T1]{fontenc}
\usepackage[hidelinks]{hyperref}
\usepackage{url}
\usepackage{booktabs}
\usepackage{amsmath}
\usepackage{amsfonts}
\usepackage{nicefrac}
\usepackage{microtype}
\usepackage{xcolor}
\usepackage{graphicx}
\usepackage{multirow}
\usepackage{subcaption}
\usepackage{enumitem}

\title{Where Does the Energy Go? Profiling LLM Agent Inference on Blackwell GPUs}

\author{%
  Qi Luo,
  Kunlin Li,
  Ziwen Wang,
  Yun Chen\thanks{Yun Chen is the corresponding author.} \\
  \normalfont Microelectronics Thrust, Function Hub, \\
  The Hong Kong University of Science and Technology (Guangzhou)
}

\begin{document}

\maketitle

\begin{abstract}
LLM agents that iteratively reason, plan, and invoke tools create workload profiles fundamentally different from single-pass inference, yet how their energy consumption is distributed across hardware components and workload phases remains poorly understood. Characterizing these workloads therefore requires simultaneous visibility into both component-level power and phase-level execution. We conduct a full-stack energy profiling study combining NVML GPU counters, Intel RAPL CPU/DRAM counters, and Intelligent Platform Management Interface (IPMI) system-level sensors on 2$\times$ NVIDIA RTX PRO 6000 Blackwell GPUs, and profile three representative workloads with Qwen3.8-27B. For mathematical reasoning, we compare thinking-enabled and thinking-disabled modes. Our measurements reveal that GPU-only telemetry misses 41--45\% of system energy across all three workloads, with non-GPU components accounting for the remainder. In our setup, the sequential agent workload consumes 63$\times$ more system energy per output token than saturated serving, reflecting the absence of batching, context growth across turns, and tool-induced idle periods. Extended thinking generates 21--75\% more tokens per problem, while per-token energy differs by less than 1\% between modes within each dataset, indicating that the resulting increase in energy is driven by output volume rather than a change in per-token efficiency. Continuous batching improves system-level energy efficiency by 3.2$\times$ from 1 to 16 requests per second, as GPU power plateaus while throughput continues to increase with batching depth. In agent and reasoning workloads, energy increases mainly because the model generates more tokens or runs longer as context grows, while system power changes little. For the memory-bandwidth-bound reasoning workload, throughput remains unchanged under per-GPU power caps of 400--600\,W but drops sharply at 300\,W. These findings suggest that context-management techniques such as summarization and selective retrieval may help reduce energy consumption in agent deployments.

\end{abstract}

\section{Introduction}

LLM agents iteratively reason, plan, and invoke tools to accomplish complex tasks~\citep{jimenez2024swebench,yang2024sweagent}. A single agentic session may span many turns, accumulate context over time, and alternate between GPU-intensive inference bursts and CPU-bound tool execution (shell commands, file I/O, and test runs). This workload profile differs fundamentally from conventional single-pass or batched inference. Reasoning-augmented models further complicate energy characterization by allocating variable compute at test time~\citep{wei2022chain,deepseekai2025r1,snell2024scaling}. How much energy does LLM inference consume across different workload regimes, and where does that energy go?

The energy cost of training has received considerable attention~\citep{strubell2019energy,patterson2021carbon}, yet inference energy remains less well characterized. At service scale, the cumulative energy demand of inference becomes substantial~\citep{luccioni2023power}. This issue is especially relevant to agent workloads in software engineering, research assistance, and autonomous decision-making. To our knowledge, prior work has not provided fine-grained, full-stack energy attribution across complete agent sessions.

Existing inference-energy studies range from GPU-only measurements~\citep{samsi2023words} to multi-level datacenter characterization using GPU, server, and row telemetry~\citep{patel2024characterizing}. GPU-only measurements omit non-GPU energy, which we show accounts for 41--45\% of total system energy. Existing datacenter studies do not capture the tool-interleaved phases of agent workloads. We provide full-stack profiling that combines GPU, CPU, DRAM, and whole-system measurements through the Intelligent Platform Management Interface (IPMI), specifically for agent workloads on Blackwell GPUs.

Profiling agent energy is challenging because the workload is non-stationary. Context grows with each turn and increases prefill duration, tool calls create gaps between model invocations, and extended thinking changes output token volume. Within each reasoning dataset, per-token energy differs by less than 1\% between thinking modes, so the total-energy difference is driven by output token volume. Across workloads, however, per-token energy reflects differences in batching, context accumulation, and tool-induced GPU idle periods. In our setup, the sequential agent workload consumes 63$\times$ more system energy per output token than saturated serving, while continuous batching improves system-level energy efficiency by 3.2$\times$ from 1 to 16 requests per second. For the reasoning workload, throughput remains unchanged under per-GPU power caps of 400--600\,W but drops sharply at 300\,W. These observations point to context and token management as promising targets for agent energy optimization.

Our work makes three contributions:

\begin{itemize}[leftmargin=*,nosep]
    \item We combine NVML, Intel RAPL, and IPMI measurements to characterize GPU, CPU, DRAM, and whole-system power. Across the workloads studied, GPU-only measurements omit 41--45\% of total system power.

    \item We show that per-turn energy generally increases with turn number while system power remains relatively stable. The increase tracks longer turn duration as context accumulates, motivating context management as a potential direction for reducing agent energy use.

    \item We find that extended thinking generates 21--75\% more tokens per problem, while per-token energy differs by less than 1\% between modes within each dataset. This indicates that the additional energy in our experiments is driven by output token volume rather than a change in per-token efficiency.
\end{itemize}

\section{Related Work}

Prior work has improved LLM inference efficiency through optimized attention kernels~\citep{dao2023flashattention2}, serving systems~\citep{kwon2023vllm,yu2022orca,stojkovic2024dynamollm}, and post-training quantization~\citep{lin2023awq}. Energy studies, however, vary in measurement scope and workload assumptions, and typically evaluate independent requests or replayed traces on pre-Blackwell hardware. Agent workloads introduce a different execution pattern because model calls alternate with tool execution and each request depends on the preceding trajectory. We organize prior work around measurement scope, serving systems and reasoning workloads, and agent workloads.

\paragraph{Measurement scope.}
Prior energy studies vary in measurement scope. \citet{samsi2023words} report aggregate GPU energy using \texttt{nvidia-smi} and DCGM. \citet{burtscher2014measuring} study the sampling behavior and distortion of an on-board GPU power sensor. \citet{luccioni2023power} extend the measurement boundary with software estimates for GPU, CPU, and memory energy under sequential inference. Recent benchmarks emphasize reproducibility across models and engines. ML.ENERGY~\citep{chung2025mlenergy} standardizes software-visible GPU energy measurement under steady-state serving. TokenPowerBench~\citep{niu2026tokenpowerbench} combines GPU telemetry, RAPL, and optional node-level instrumentation, and separates prefill energy from decode energy. JouleShare~\citep{luo2026requestlevel} attributes aggregate GPU energy to individual requests in static and continuous batches using replay-based Shapley ground truth. Together, these studies characterize energy use at the accelerator, component, and request levels, but they evaluate predetermined requests whose contents do not depend on prior tool interactions. They therefore do not capture complete tool-interleaved trajectories or the system energy consumed between an agent's model calls.

\paragraph{Serving systems and reasoning workloads.}
Serving systems improve LLM inference efficiency in different ways. DynamoLLM~\citep{stojkovic2024dynamollm} adjusts GPU frequency and tensor parallelism under latency constraints, while DistServe~\citep{zhong2024distserve} separates prefill from decoding. These systems improve serving efficiency but do not characterize full-system energy at the request level. Separate work examines the energy implications of reasoning. Reasoning methods can increase inference work through longer reasoning traces or multiple sampled solutions~\citep{snell2024scaling,wei2022chain,deepseekai2025r1}. \citet{jin2025energyreasoning} show that majority voting and reasoning-token scaling lead to task- and model-dependent energy increases with diminishing accuracy returns. Both lines of work evaluate controlled queries or replayed traces rather than tool-interleaved agent workloads, where tool output determines the next model request and non-model intervals contribute to total task energy.

\paragraph{Agent workloads.}
Agent workloads introduce trajectory dependencies that conventional serving benchmarks do not capture. ReAct~\citep{yao2023react} and SWE-agent~\citep{yang2024sweagent} execute multi-step tasks in which later actions depend on tool feedback. AgentStop~\citep{pham2026agentstop} is the closest energy-focused study. It uses early termination to reduce energy waste from failed runs and measures CPU and GPU components on an Apple M1 Max. Our study profiles complete agent trajectories on Blackwell GPUs~\citep{nvidia2025blackwell} using both GPU-level and whole-system measurements. To our knowledge, prior work has not evaluated tool-interleaved agent trajectories with both measurement scopes alongside a continuous-serving baseline on the same platform.

\section{Methodology}

In this section, we describe the serving configuration, measurement interfaces, workloads, and evaluation metrics used in our experiments.

\begin{figure}[t]
    \centering
    \includegraphics[width=0.7\linewidth]{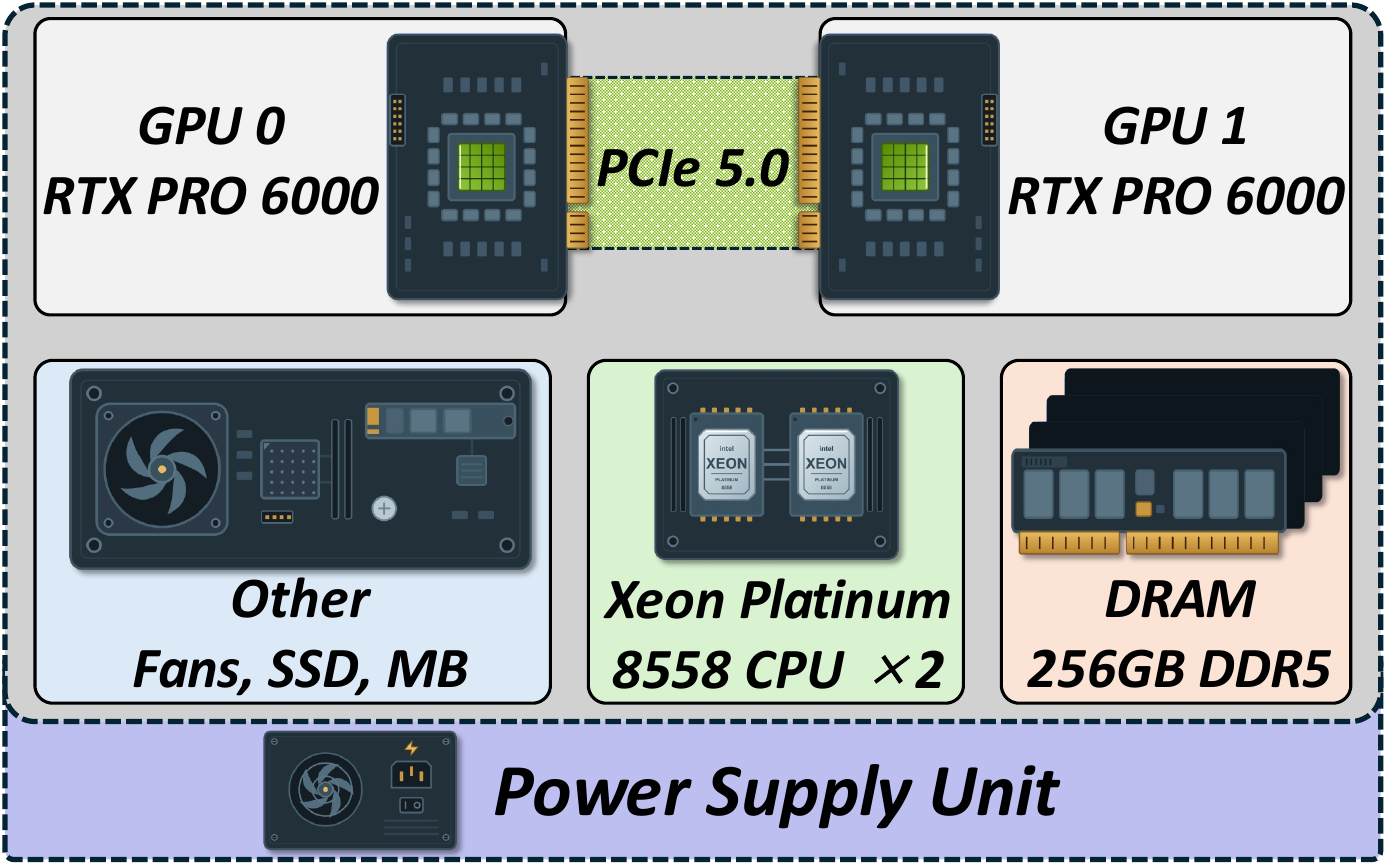}
    \caption{System power measurement architecture on the dual-GPU server.}
    \label{fig:system_arch}
\end{figure}

\subsection{Platform and Serving}

All experiments run on a GPU server equipped with:

\begin{itemize}[leftmargin=*,nosep]
    \item \textbf{GPU:} 2$\times$ NVIDIA RTX PRO 6000 Blackwell Server Edition GPUs (96\,GB GDDR7 each, configurable up to 600\,W), each connected to the host via PCIe 5.0 $\times$16.
    \item \textbf{CPU:} 2$\times$ Intel Xeon Platinum 8558 (48 cores/96 threads, 330\,W TDP each).
    \item \textbf{Memory:} 256\,GB DDR5 ECC.
    \item \textbf{Software:} Ubuntu 26.04, CUDA 13.3, Driver 610.57.
\end{itemize}
The server's baseboard management controller (BMC) exposes IPMI sensors for system-level power measurement. We serve Qwen3.8-27B~\citep{qwen38} via vLLM~\citep{kwon2023vllm}, version 0.19.1, with tensor parallelism across both GPUs, using BF16 precision, chunked prefill, and a maximum context length of 131k tokens.

\subsection{Energy Profiling}

As shown in Figure~\ref{fig:system_arch}, we collect power and energy measurements from three interfaces within a common timing loop:

\begin{itemize}[leftmargin=*,nosep]
    \item \textbf{NVML:} Cumulative per-GPU energy counters sampled at 10\,Hz using \texttt{nvmlDeviceGetTotalEnergyConsumption}.
    \item \textbf{RAPL}~\citep{khan2018rapl}: CPU package and DRAM energy counters sampled at 10\,Hz through \texttt{/sys/class/powercap}.
    \item \textbf{IPMI:} CPU, memory, fan, and total-system power readings sampled at 1\,Hz from BMC sensors.
\end{itemize}

We assess cross-interface consistency by comparing RAPL CPU package power with IPMI CPU rail readings. Across the evaluated workloads, the two measurements differ by only 0.1--0.6\%.

IPMI reports CPU, memory, and fan power separately, along with total system power. We define $P_\text{others} = P_\text{system} - P_\text{GPU} - P_\text{CPU} - P_\text{DRAM}$. This residual includes contributions from fans, voltage regulator module (VRM) conversion losses, power supply unit (PSU) overhead, the baseboard, SSDs, the BMC, and other components not measured separately.

\subsection{Workloads}

We evaluate the following three types of workloads:

\begin{itemize}[leftmargin=*,nosep]
    \item \textbf{Agentic coding.} We run SWE-bench Verified~\citep{chowdhury2024swebenchverified} tasks in official Docker containers using multi-turn tool calling and extended thinking.
    \item \textbf{Mathematical reasoning.} We evaluate all 90 problems from AIME 2022--2024~\citep{aime} and MATH-500~\citep{lightman2023lets} with thinking enabled and disabled to isolate the energy overhead of extended thinking.
    \item \textbf{Continuous serving.} We serve ShareGPT~\citep{zheng2024judging} prompts at arrival rates of 1--16 requests per second under continuous batching to characterize the throughput-efficiency frontier.
\end{itemize}

\subsection{Metrics}

We report the following energy and power metrics:

\begin{itemize}[leftmargin=*,nosep]
    \item \textbf{Energy per token} (mJ/tok): total system energy divided by output tokens.
    \item \textbf{Power} (W): power measured at the component and whole-system levels.
    \item \textbf{Energy per task} (kJ): total system energy to complete one problem or request.
\end{itemize}
Energy metrics use total system energy unless explicitly stated as GPU-only. Power is reported at both component and system levels.

\section{Experiments}

In this section, we present results for the three workload types, compare their energy profiles, and evaluate the effects of GPU power capping.

\subsection{Agentic Coding}

\begin{figure}[t]
    \centering
    \includegraphics[width=0.95\textwidth]{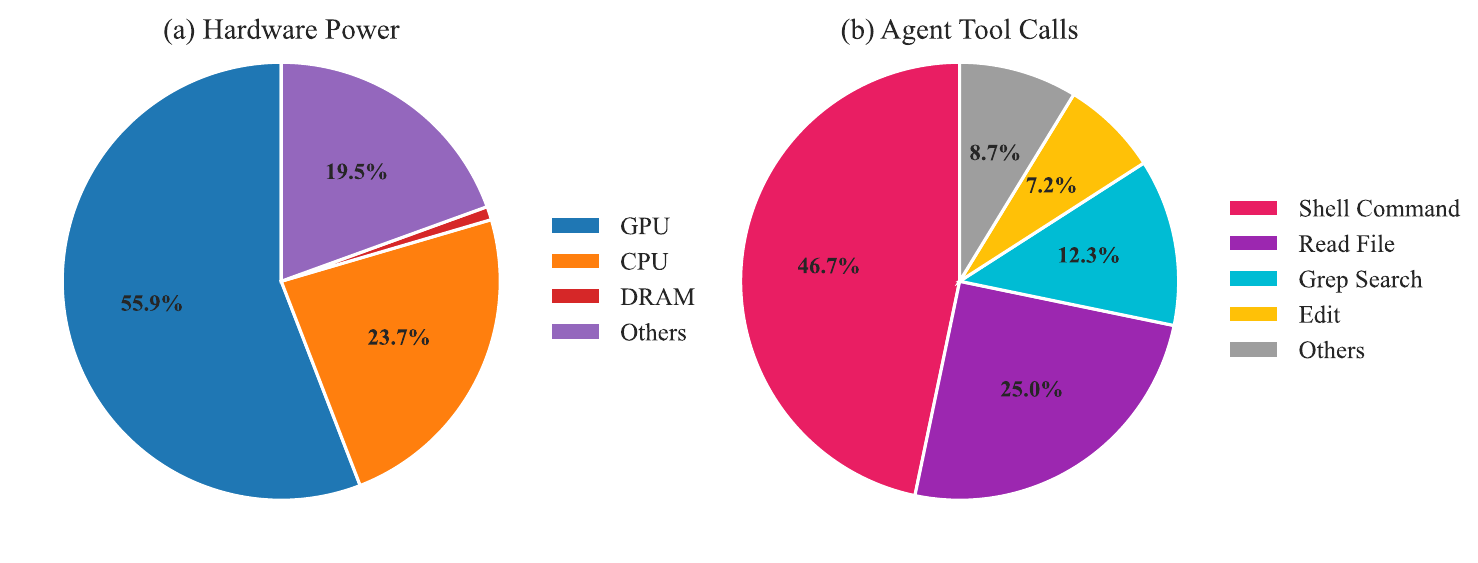}
    \caption{SWE-bench agentic coding workload: (a) system power decomposition, (b) agent tool call distribution.}
    \label{fig:decomposition}
\end{figure}

We run SWE-bench Verified tasks using the official Docker harness orchestrated by Harbor with a Qwen3.8-27B agent in single-agent sequential mode. Due to computational resource constraints, we limit our evaluation to 50 tasks. Figure~\ref{fig:decomposition} summarizes both hardware power distribution and agent behavior.

\paragraph{System power decomposition.}
Total system power averages 1,406\,W during agentic coding. The two GPUs account for the largest share at 55.9\% (785\,W combined), followed by the CPU at 23.7\% (333\,W across two sockets). DRAM contributes only 1.0\% (14\,W), consistent with the model residing entirely in GPU memory. The remaining 19.5\% (274\,W) is categorized as ``Others.'' This residual includes estimated contributions from cooling fans ($\sim$133\,W), VRM conversion losses ($\sim$50\,W at 92--95\% conversion efficiency), PSU AC--DC conversion overhead ($\sim$50\,W at 80 Plus Platinum rating), PCIe link power ($\sim$10\,W for the two $\times$16 links), and baseboard components, including SSDs and the BMC ($\sim$31\,W). GPU-only telemetry therefore misses approximately 44\% of total system energy.

\paragraph{Agent behavior.}
We analyze the distribution of agent tool calls to characterize agent activity. Shell commands account for the largest share at 46.7\%, followed by file reads (25.0\%), grep searches (12.3\%), and edits (7.2\%). Shell commands often launch subprocesses or run tests while the GPU remains largely idle and the CPU stays active, contributing to the sustained CPU power draw between inference bursts.

\begin{figure}[t]
    \centering
    \includegraphics[width=0.95\textwidth]{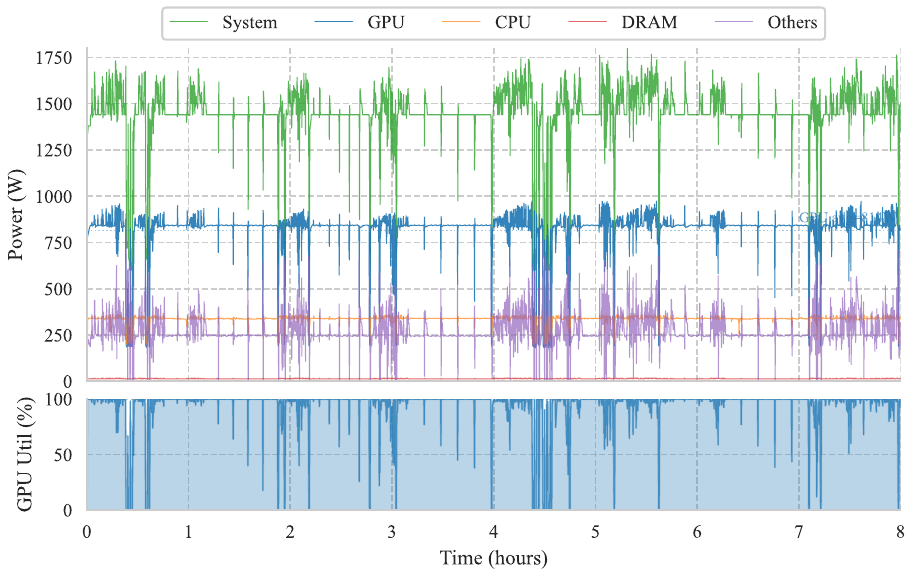}
    \caption{Agent power trace over a selected 8\,h window of SWE-bench execution. Top: component-level power. Bottom: GPU utilization.}
    \label{fig:power_trace}
\end{figure}

\paragraph{Power trace reveals burst/idle pattern.}
Figure~\ref{fig:power_trace} shows the temporal power profile over a selected 8-hour window of an agent session. GPU power follows a burst/idle pattern, rising to $\sim$800\,W during inference and falling to $\sim$150\,W during tool execution. The bottom panel shows corresponding changes in GPU utilization, from near-100\% during generation to near-0\% during tool calls. System power generally remains elevated between inference bursts because the CPU stays active during operations such as shell commands and test execution.

\begin{figure}[t]
    \centering
    \includegraphics[width=0.95\textwidth]{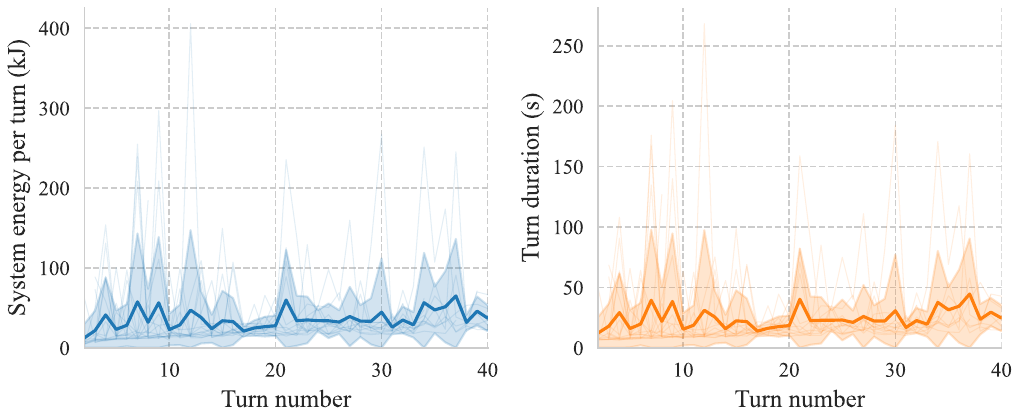}
    \caption{Per-turn system energy and duration across SWE-bench tasks. Light lines: individual tasks; bold: mean $\pm$1$\sigma$.}
    \label{fig:perturn}
\end{figure}

\paragraph{Per-turn energy tracks execution duration.}
Figure~\ref{fig:perturn} shows that both system energy and duration per turn tend to increase over the course of an agent trajectory. Average system power changes little across turns, so the similar trends in energy and duration indicate that longer execution is the primary source of the energy increase. This pattern is consistent with longer prefill as context accumulates across turns.

\subsection{Mathematical Reasoning}

We evaluate the model on AIME and MATH-500 with thinking enabled and disabled. As shown in Table~\ref{tab:ept}, per-token energy differs by less than 0.4\% between modes at both the component and system levels. GPU per-token energy is 17,531 vs.\ 17,501\,mJ/tok on AIME and 17,505 vs.\ 17,449\,mJ/tok on MATH-500. Across these settings, the GPU accounts for 58.8\% of system energy, while the CPU accounts for approximately 23.5\%. Thinking-enabled runs generate 75\% more tokens per problem on AIME and 21\% more on MATH-500, with corresponding increases in total energy per problem. This indicates that the additional energy primarily reflects greater output volume rather than a change in per-token efficiency. AIME problems are harder and elicit longer reasoning chains, resulting in higher energy consumption per problem than MATH-500.

\begin{table}[ht]
\centering
\caption{Mathematical reasoning energy profile. $\checkmark$ = thinking enabled, $\times$ = thinking disabled.}
\label{tab:ept}
\begin{tabular}{llccccccc}
\toprule
& & \multicolumn{5}{c}{\textbf{Energy per Token (mJ/tok)}} & \multicolumn{2}{c}{\textbf{Per Problem}} \\
\cmidrule(lr){3-7} \cmidrule(lr){8-9}
\multirow{-2}{*}{\textbf{Dataset}} & & \textbf{GPU} & \textbf{CPU} & \textbf{DRAM} & \textbf{Others} & \textbf{System} & \textbf{Tokens} & \textbf{Energy (kJ)} \\
\midrule
AIME & \checkmark & 17,531 & 7,018 & 268 & 4,985 & 29,802 & 10,506 & 313.2 \\
AIME & $\times$ & 17,501 & 7,000 & 268 & 4,982 & 29,751 & 5,991 & 178.2 \\
\midrule
MATH-500 & \checkmark & 17,505 & 7,008 & 268 & 4,978 & 29,759 & 1,800 & 53.6 \\
MATH-500 & $\times$ & 17,449 & 7,004 & 267 & 4,962 & 29,682 & 1,488 & 44.2 \\
\bottomrule
\end{tabular}
\end{table}

\subsection{Continuous Serving}

We serve ShareGPT conversation prompts using vLLM with continuous batching at request arrival rates of 1, 2, 4, 8, and 16 requests per second. Table~\ref{tab:serving} summarizes the resulting power, throughput, and energy efficiency. Unlike agentic coding and mathematical reasoning, which process one model request at a time, continuous serving batches multiple in-flight requests, with up to approximately 20 requests decoded in parallel at higher loads. At 16 requests per second, aggregate throughput reaches 1,494 tokens per second, while GPU power rises from 662\,W at 1 request per second to 802\,W. Over the same range, system energy per token falls from 2,955 to 920\,mJ/tok, corresponding to a 3.2$\times$ improvement in energy efficiency. GPU power largely plateaus by 4 requests per second, while throughput continues to increase at higher request rates with relatively little additional GPU power. Figure~\ref{fig:efficiency_frontier} visualizes this throughput-efficiency frontier.

\begin{table}[ht]
\centering
\caption{Continuous serving energy profile at increasing request rates.}
\label{tab:serving}
\begin{tabular}{cccccccc}
\toprule
& \multicolumn{5}{c}{\textbf{Power (W)}} & & \\
\cmidrule(lr){2-6}
\multirow{-2}{*}{\textbf{req/s}} & \textbf{GPU} & \textbf{CPU} & \textbf{DRAM} & \textbf{Others} & \textbf{System} & \multirow{-2}{*}{\shortstack{\textbf{Throughput}\\(tok/s)}} & \multirow{-2}{*}{\shortstack{\textbf{System}\\(mJ/tok)}} \\
\midrule
1  & 662  & 342 & 13 & 169 & 1,185 & 401   & 2,955 \\
2  & 756  & 340 & 13 & 156 & 1,265 & 595   & 2,125 \\
4  & 792  & 341 & 13 & 181 & 1,327 & 1,158 & 1,147 \\
8  & 800  & 340 & 13 & 201 & 1,354 & 1,449 & 935 \\
16 & 802  & 340 & 13 & 219 & 1,375 & 1,494 & 920 \\
\bottomrule
\end{tabular}
\end{table}

\subsection{Cross-Workload Comparison}

\begin{figure}[t]
    \centering
    \includegraphics[width=0.48\textwidth]{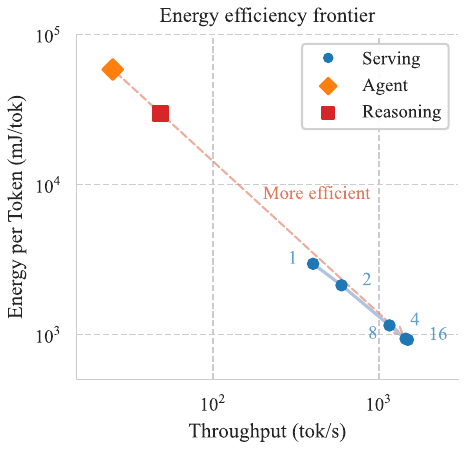}
    \caption{Energy efficiency frontier across workloads (log-log scale).}
    \label{fig:efficiency_frontier}
\end{figure}

Figure~\ref{fig:efficiency_frontier} compares energy efficiency across workloads. At the system level, the agent workload consumes 58,383\,mJ per output token, approximately 63$\times$ that of saturated serving at 16 requests per second (920\,mJ/tok). Three workload characteristics help explain this gap: (1)~single-agent execution provides no opportunity for batching across requests, (2)~context growth increases prefill work across turns, and (3)~interleaving tool execution with model calls adds non-inference time and energy to the agent trajectory. In absolute terms, completing one SWE-bench task costs approximately 2,989\,kJ, compared with 313.2\,kJ and 53.6\,kJ for thinking-enabled AIME and MATH-500 problems, respectively.

The GPU-to-system power ratio remains within 55--59\% across the workload types. \citet{patel2024characterizing} report a roughly 60\% GPU-to-server power ratio for LLM inference. Our measurements complement this aggregate ratio with a finer-grained component-level breakdown across agentic coding, mathematical reasoning, and continuous serving. Our slightly lower range is consistent with the two high-TDP CPUs in our platform (2$\times$330\,W Xeons paired with 2$\times$600\,W GPUs). Reporting only GPU energy would omit the remaining 41--45\% contributed by the CPU, DRAM, and other components. The relatively narrow range suggests that this gap is strongly influenced by platform configuration, although its exact value remains workload dependent.

\subsection{Power Capping}

\begin{figure}[t]
    \centering
    \includegraphics[width=0.95\textwidth]{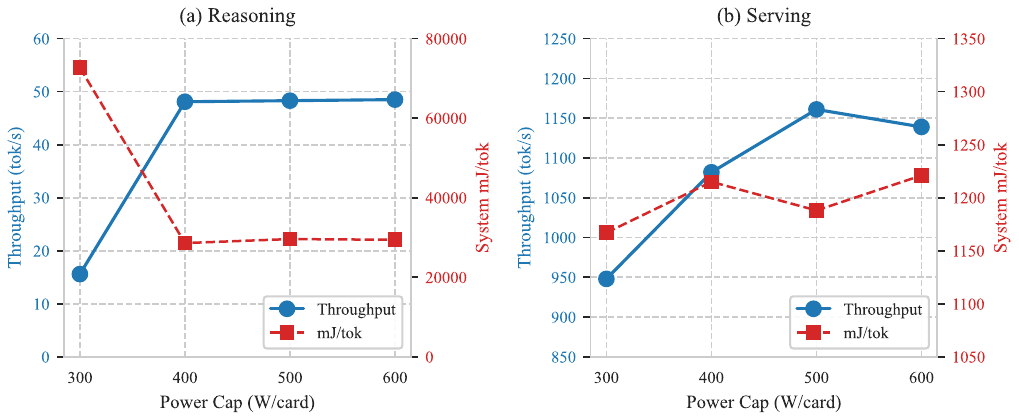}
    \caption{Power cap sweep. (a) Reasoning, single request. (b) Serving, 4 requests per second. Both panels show throughput (left axis) and system energy per token (right axis).}
    \label{fig:powercap}
\end{figure}

We sweep per-GPU power limits from 300 to 600\,W in 100\,W steps. Figure~\ref{fig:powercap} summarizes the resulting throughput and system energy per token for reasoning and serving. For reasoning, throughput remains approximately 48 tokens per second under power caps of 400--600\,W. The 500\,W and 600\,W caps are non-binding because actual power draw is approximately 420\,W per card, while the 400\,W cap reduces GPU power without affecting throughput, consistent with memory-bandwidth-bound decoding. At 300\,W, clock throttling reduces throughput by a factor of 3.1 and raises system energy per token to 2.5$\times$ its value at 400\,W, producing a sharp performance cliff. The combined power draw of non-GPU components varies only modestly across the cap settings compared with GPU power, indicating that the system-level response to power capping is concentrated primarily in the GPUs. For serving at 4 requests per second, the 300\,W cap achieves the lowest system energy per token (1,167 vs.\ 1,215\,mJ/tok at 400\,W) because the reduction in power outweighs the loss in throughput.

\section{Discussion}

In our setup, the sequential agent workload uses approximately 63$\times$ as much system energy per output token as saturated serving. Within agent trajectories, per-turn energy tracks execution duration rather than average power, consistent with longer prefill as context grows. This suggests that context management may help reduce agent energy use.

Within each dataset, enabling thinking changes per-token energy by less than 1\% but increases output tokens per problem by 21--75\%. The added energy therefore mainly reflects token volume. Early stopping and budget-aware decoding may reduce this overhead, subject to accuracy tradeoffs.

In our setup, GPU telemetry captures 55--59\% of total system energy, leaving 41--45\% unreported. The limited variation across workload types suggests that this gap is strongly shaped by platform configuration, although the exact share remains workload dependent. Carbon estimates should therefore use full-system measurements or platform-specific corrections.

Continuous batching improves energy efficiency by 3.2$\times$ from 1 to 16 requests per second as GPU power largely plateaus. For reasoning, the 500\,W and 600\,W caps are non-binding, the 400\,W cap reduces GPU power without affecting throughput, and the 300\,W cap causes a sharp performance drop. This pattern highlights the need for workload-specific power-cap selection.

\paragraph{Limitations.}
We study one dual-GPU server and one 27B model, so generalization to other platforms and scales remains uncertain. We lack external power distribution unit (PDU) measurements for AC validation, and the ``Others'' category includes estimated contributions.

\section{Conclusion}

We present a full-stack energy profiling study of agentic coding, mathematical reasoning, and continuous serving on Blackwell GPUs. The results show how context growth, output-token volume, batching, and power limits shape energy use across these workload types. They also show that GPU telemetry alone is insufficient for full-system energy accounting.

\bibliographystyle{plainnat}
\bibliography{references}

\end{document}